\documentclass[twocolumn,twoside,preprintnumbers,amsmath,amssymb,showkeys]{revtex4}

\usepackage{graphicx}

\usepackage{fancyhdr}
\usepackage{pslatex}

\begin{document}

\title {Status and Promise of Particle Interferometry in Heavy-Ion Collisions}

\author{Selemon Bekele, Fabio Braghin, Zbigniew Chaj\c ecki , Paul Chung, John G. Cramer,
Tam\'as Cs\"org\H o, Hans Eggers, Sean Gavin, Fr\'ed\'erique
Grassi, Yogiro Hama, Adam Kisiel, Che-Ming Ko, Tomoi Koide,
Gast\~ao Krein, Roy Lacey, Richard Lednicky, Michael A. Lisa,
Wesley Metzger, Dariusz Mi\'skowiec, Kenji Morita, Sandra S.
Padula, Scott Pratt, Wei-Liang Qian, Vladislav \v{S}imak, Yuri
Sinyukov, Michal \v{S}umbera, Bernardo M. Tavares, Giuseppe Verde,
Detlef Zschiesche}

\vskip0.5cm \affiliation{\vskip0.3cm Participants of the II 
Workshop on Particle Correlations and Femtoscopy (WPCF 2006) \\
Instituto de F\'\i sica Te\'orica - UNESP \\
S\~ao Paulo, Brazil, September 9, 10 \& 11, 2006 }
%

\begin{abstract}
After five years of running at RHIC, and on the eve of the LHC
heavy-ion program, we highlight 
the status of femtoscopic
measurements. We emphasize the role interferometry plays in
addressing fundamental questions about the state of matter created
in such collisions, and present an enumerated list of
measurements, analyses and calculations that are needed to advance
the field in the coming years.
\end{abstract}

\vskip -1.35cm

\maketitle

\thispagestyle{fancy}

\setcounter{page}{1}

\bigskip


The first two-pion interferometric measurements of the collider
era emerged within a year of RHIC becoming operational. To the
surprise of many in the field, the measurements were remarkably
similar to those recorded at the AGS and the SPS. The analyses
were inconsistent, both qualitatively and quantitatively, with
dynamic models incorporating first-order phase transitions in
general. 
In fact, parametric explanations of the data, the forms of which
are often motivated by solutions to dynamical equations, suggest
that the matter explodes violently, growing from a radius of 6 fm
to 13 fm in only 10 fm/c. The surprisingly strong acceleration
required for such behavior and the associated failure of many of
the field's most sophisticated models became known as the ``HBT
Puzzle''.

Five years later, the field has made steady progress on a variety of fronts.
Theoretically, sophisticated dynamic models have more successfully
reproduced experimental results (though the very sophistication of
these models has made it difficult to ascertain which aspects of
the models are being validated by the comparison). New techniques
have been applied to the analysis of experimental correlation
functions, revealing greater detail about the size and shape of
the emission region. During our discussions at the workshop, it
was clear that a remarkable consensus had developed among the
practitioners of the field. Although this agreement by no means
represented a final conclusion, we found numerous points that
could be stated without dissent. In this white paper we will first
list the points concerning the current status of the field, then
further below, enumerate points where participants agreed were
important
for further progress.\\

\noindent {\large\bf Achievements:}\vspace*{-4pt} \medskip
\begin{itemize}

\item Remarkable agreement has been observed between the RHIC
experiments, PHENIX, PHOBOS and STAR. All three have produced
high-statistics high-quality pion correlations, whose apparent
source sizes are consistent to a few tenths of a fm. A similar
consistency was observed among measurements performed at the top
SPS energy; at lower SPS energies the maximum deviations are on
the level of 20\%.

\item Femtoscopic studies are highly multi-dimensional. Even the
simplest and most common case of two-identical pion correlations
depend on six independent variables, which have only been fully
explored within the past few years. 
This includes extracting characteristic source sizes as function
of transverse momentum, rapidity and the angle with respect to the
reaction plane for off-axis collisions. Additionally, correlation
functions have been analyzed as a function of beam energy,
centrality and the species of the pair. This accomplishment is
especially noteworthy when taking the perspective of comparing to
the field 20 years ago, when extracting a single source dimension
was considered state-of-the-art.

\item The majority of femtoscopic investigations continues to
focus on the correlations of identical pions, but analyses
involving numerous other pairs (even $\Xi-\pi$ correlations) are
becoming more common. Analysis of non-identical particle
correlations has allowed the extraction of qualitatively new
femtoscopic information about the dynamical source substructure.
Thus far, the preliminary assessment is that they are consistent
with the information gleaned from $\pi-\pi$ correlations.
Particularly, the measured pion-proton and pion-kaon correlation
asymmetries point rather directly to a strong collective flow in
heavy ion collisions at SPS and RHIC.

\item Advanced techniques for angular decomposition and imaging
are now being applied to extract shape and size information from
any measured correlation. Although these analyses are in their
nascent stage, it appears they are uncovering quantitative details
about the longer-time-scale aspects of particle emission, such as
resonance production or surface emission.

\item Without doubt, {\it the} dominant dynamical feature of the
bulk system created at RHIC is its explosive collective motion
(flow). Flow generates a source with characteristic
dynamical/geometric substructure, and has implications (e.g.
spectral shapes and ``$v_2$'' anisotropy) when projected onto the
momentum-only space.  However, with its explicit focus on the
space-momentum source substructure, interferometry is the most
sensitive and detailed probe of collective flow. The growth of the
sideward and longitudinal sizes, along with the lack of
significant extension of the outward direction, related to a short
duration of the particle emission, can only be reproduced with
highly explosive dynamics. Parametric descriptions based on
thermal emission on the background of large collective outward
flow also explain the large radial shift in emission points of
different-mass particles and the dependence of the effective
source sizes with transverse momentum and direction with respect
to the reaction plane. Measurements of effective sizes and
orientation of the source shape for non-central rapidities have
also validated our space-time picture of longitudinal collective
flow. Analyses with a wide range of models always come to the same
conclusion, qualitatively and quantitatively, that strong
longitudinal and transverse flow has developed in central
collisions at the SPS and at RHIC.

\item The relative success and failure of various dynamic
descriptions to provide sources that match those observed with
interferometry has significantly constrained our understanding of
the equation of state at high temperature. Twenty years ago,
bag-model descriptions of the equation of state with latent heats
of many GeV/fm$^3$ were common. It is now clear that extremely
soft equations of state, i.e., those that have large latent heats,
are grossly inconsistent with interferometric measurements.
Although smaller latent heats are not yet ruled out
(
cross-over or second order transitions are also possible), the
range of acceptable equations of state would be much broader if
not for femtoscopic analyses.

\item The pion source sizes, when analyzed as a function of the
collision energy, seem to follow the mean pion cross section for
scattering on surrounding particles in the collision fireball.
This indicates that the freeze-out happens when the pion mean free
path exceeds a certain critical value, in a quantitative analysis
estimated to be 1-2 fm. The interferometry data favors the fixed
mean free path freeze-out criterion over a freeze-out at a fixed
spatial density, or at a fixed phase-space density, or when the
mean free path exceeds the system size.

 \item In addition to hints about the equation of state
that would manifest themselves through dynamics, and thus through
observables such as $R_{\rm out}/R_{\rm side}$, correlations have
provided a quasi-model-independent measure of the phase space
density and the total entropy observed in heavy ion collisions.
Although the first estimates are rather rough, this already has
provided a significant constraint on the equation of state.

\item The connection between the source emission probability,
which is given in coordinate space, and the correlation function,
measured as a function of relative momentum, is contingent on the
assumption of chaotic uncorrelated emission sources, whose
correlations arise principally from final-state two-body
interactions. This presumption has been verified by measurements
of three-pion correlations, which have been shown theoretically to
be sensitive to coherent emission.

\item  A parallel direction has developed, distinct from the study
of heavy ion collisions, {\it per se.} Extracted knowledge of the
space-time substructure of the emitting source also allows the
femtoscopic program to be run ``in reverse.'' Assuming known
geometries from other correlations analyses, correlations for
pairs where the interactions are not well understood (e.g.
$\Lambda\Lambda$) are being used to determine details of the
interaction between unstable particles. Since correlation analyses naturally involve low-relative momentum pairs, scattering lengths are especially accessible.

\end{itemize}

Despite the enormous progress listed above, significant hurdles
have yet to be overcome and numerous opportunities have not yet
been exploited. Workshop participants felt that enumeration of a
``to-do'' list for the field would be enormously helpful, both for
informing the greater heavy-ion community of our plans, and for
clarifying, in our own minds, the important needs for our
immediate future. These needs encompass both new experimental
equipment, measurements, and analyses, along with needed
development in theory, and with better integrating interferometric
analyses with other families of
observables.\\

\noindent {\bf\large Opportunities and Challenges:}\vspace*{-4pt}
\medskip
\begin{itemize}

\item Coherent phenomena are intimately associated with
correlations. This includes novel Bose effects as well as coherent
emission from classical fields, such as what is often described in
dynamic models of the chiral condensate. In nearly all such cases,
the phenomena are expected to be strongest at low $\mathbf{p_T}$.
Since the momentum scale can be estimated as the inverse
characteristic source size, measurements at $\mathbf{p_T}\sim 50$
MeV/$c$ are required to best explore such possibilities. Such
measurements might require experiments to either run at low
magnetic field settings or to install special detectors.

\item Back-to-back correlations (BBC) have recently been shown to
arise if hadronic masses are modified by interactions in a
dense medium. These quantum mechanical correlations are induced by
a non-zero overlap between the in-medium states and free states,
which are observed. In particular, medium-modified bosonic or
fermionic fields can be represented in terms of two-mode squeezed
states of the corresponding asymptotic fields. Both the fermionic
and the bosonic BBC lead to positive correlations of unlimited
strength. They are more pronounced for large absolute values of
the particles' back-to-back momenta and might survive the
effect of collective flow. A joint experimental and
theoretical effort will be essential for effectively observing and
understanding such correlations, by looking for both an optimized
form of the signal and the most promising experimental conditions
for measuring it.

\item Interferometric and flow analyses at RHIC have suggested
that the average speed of sound 
is in the neighborhood of $c_s \sim 0.3 - 0.4$. It is expected
that the LHC region will explore much further above $T_c$ where
the matter is predicted to stiffen  and the speed of sound
approaches $c_s \sim 1/\sqrt{3} \approx 0.58 $. It is imperative
that experiments at the LHC have the capability for making high
quality correlation measurements for particles with $100$ MeV/$c
\lesssim \mathbf{p_T} \lesssim 1$GeV/$c$ if this fundamental
property of hot matter is to be explored.

\item In the last few years, lattice calculations have begun to
suggest that the QCD phase transition is a cross-over for nearly
zero chemical potentials, probed at RHIC with $\sqrt{s_{NN}} =
200$ GeV collisions, and that it becomes a second order phase
transition at a critical end point in the phase-diagram
($T_{_{CEP}}, \mu_{_{CEP}}$). Beyond this critical value, and for
higher chemical potentials, $\mu > \mu_{_{CEP}}$, the transition
becomes first order. The critical point might be reached for
energies just above the AGS range, in low energy runs at the CERN
SPS and at the RHIC accelerators. As this range will also be
covered by the upcoming FAIR facility at GSI and at the planned NICA
facility at JINR, it is important that proper detectors are
installed for high-resolution measurements of correlations in the
critical region.

\item An analysis of the energy dependence of the averaged
particle phase-space density, which is directly related to
femtoscopy measurements in current and future experiments, is  of
great interest. This quantity is approximately conserved during
the hadronic stage of evolution and therefore is connected with
the initial phase-space density of hadronic matter. It provides
information about the states of the matter at the end of
hadronization stage, or at chemical freeze-out, and thus allows one to
search for phase transitions or a limiting Hagedorn temperature in
relativistic nucleus-nucleus collisions.

\item Femtoscopy for high $\mathbf{p_T}$ particles would provide
stringent tests of theoretical pictures of recombination and
fragmentation models such as coalescence. However, femtoscopic
measurements have thus far only been made for pairs with low
relative momentum, as it is inherently difficult to gather
statistics at high $\mathbf{p_T}$,  where the phase space density
is low. Specialized measurements or intensity upgrades might be
required for high-quality interferometric measurements in the
relevant $\mathbf{p_T}$ range of several GeV/$c$.

\item Analyses of correlations for pairs of particles other
than identical pions, such as $pp$, $pK$, $\pi p$,
$p\Lambda\cdots$, are in a nascent stage, with new imaging
techniques having been recently developed for extracting detailed
size and shape information. It would be of tremendous importance
to quantitatively verify, using different classes of final-state
interactions, the space-time picture of the breakup stage that has
emerged from the analysis of identical-pion correlations. Such
analyses might require experimental upgrades, such as the STAR
time-of-flight wall, which by expanding the range of particle
identification would allow particles of much different mass to be
correlated at low relative velocity.

\item The interferometry of penetrating probes has just recently
become possible. In addition to potentially providing space-time
information about  pre-breakup stages, $\gamma\gamma$ correlations
can reveal the fraction of low $\mathbf{p_T}$ photons that do not
originate from $\pi^0$ decays, thus providing a direct photon
spectrum at sufficiently low $\mathbf{p_T}$ to yield robust
insight into the temperature during earlier stages of heavy-ion
collisions. Also, recently there has been some theoretical
development of correlation functions for lepton pairs, which
reiterates the need for measuring correlation data of other
penetrating probes, such as lepton-lepton femtoscopy.

\item Meticulous experimental analysis of the structure of the
correlation function at very small relative momentum appears to
show details of longer-time components of the emission function.
Given the connection between long-lived emission and the equation
of state, it is important to vigorously pursue such analyses. In
some cases, this might require detector upgrades to achieve the
$\sim 2$ MeV/$c$ resolution necessary to resolve low relative
momentum features in the correlation function.

\item The primary motivation of heavy ion experiments is the
determination of bulk properties of matter. To achieve this
ultimate goal, all relevant observables, including flow and
spectra, must be simultaneously analyzed by comparing
comprehensive dynamical models with data. Analyses of parametric
models have already illustrated the importance of a coordinated
study of both correlations and spectra. Although numerous dynamic
descriptions have been tested for their interferometric
predictions, many models remain untested, or are only tested
through rather primitive breakup criteria. The femtoscopic
community needs to improve our link with theorists developing
dynamical models such as hydrodynamics. Furthermore, the theory
community should be strongly encouraged to develop models which
are better tested, better documented and are more flexible.  In
particular, femtoscopic conclusions about the equation of state
(EoS) have been complicated by the need, in many cases, to compare
to the data one group's calculation using a given EoS, and {\it
another} group's calculation using a different EoS. Firmer
conclusions about this crucial feature of the matter will be much
helped if all groups will produce predictions using a variety of
EoS. Similar treatment should be adopted with other important
factors, such as initial conditions, the list of free parameters
considered in each model, etc. Such a development is crucial if
interferometric data are to be fully exploited.

\item Extracting source functions from hydrodynamic models should
be better 
accommodated, which might entail providing interfaces between
hydrodynamic models and microscopic hadronic codes used for
modelling the final breakup. Given the possible importance of mean
fields in driving the dynamics or in refracting outgoing
trajectories, mean fields should be included in transport codes.
Such codes should be made available to the hydrodynamic community,
along with support for interfacing the descriptions.

\item Given the increasing sophistication and subtlety of femtoscopic
studies, some collaboration between experimentalists and theorists
{\it on the data analysis itself} might be beneficial. Experimental
collaborations may
consider incorporating more flexibility
into rules related to propriety of data, either in general or on a case-by-case basis.
%

\item A large fraction of practicing femtoscopists, now focusing
on relativistic heavy ion collisions, initially worked at lower
(sub-AGS) energies.  Frequent interactions and exchanges of ideas
and techniques has always benefitted both communities. However,
these interactions may be becoming less frequent due to two
factors: (i) the increasing fraction of young people who already
began their career at the highest energies, and (ii) the
increasingly self-referential nature of relativistic heavy ion
physics in general.  Continued and enhanced collaboration between
high- and lower-energy femtoscopy should be an explicit
consideration in the organization of femtoscopy-oriented symposia
and workshops.

\item The heavy-ion and the high-energy femtoscopic communities
should make greater efforts towards communicating. Advancing the
understanding of small source ($\sim 1$ fm) interferometry will
require more careful analysis of numerous effects which challenge
the assumption of chaotic independent emission. In particular,
common research projects between heavy ion physicists and experts
working on interferometry studies in elementary particle
collisions, such as in $p p$, $h p$, $\bar{p}p$, $e^+ e^-$, should
be strongly encouraged. By better understanding the $\mathbf{p_T}$
dependence of source sizes in $pp$ collisions, we should attain a
quantitative understanding of the effects in $AA$ collisions and
provide a systematic error to the underlying theory.
\end{itemize}

A unifying theme of all the points, both in the list of
accomplishments and in the list of upcoming challenges, is the
importance of collaboration. This includes sharing knowledge,
expertise and ideas between collaborations, between
experimentalists and theorists, between various segments of the
theoretical community, and between different fields. To that end,
there was unanimous consent that the WPCF series of workshops has
already been enormously useful. Evidence of discussions and
collaborations during the 2005 meeting in Krom\v{e}\v{r}\'\i \v{z}
was already evident in the results shown in 2006. The continuation
of the workshop was enthusiastically endorsed by all participants.

\noindent\vskip0.5cm

\section*{Acknowledgments}
The authors gratefully acknowledge the hospitality of the
Instituto de F\'isica Te\'orica - UNESP, in S\~ao Paulo, for so
graciously providing such a pleasant and stimulating environment
for the workshop.

The organizers would like 
to thank CAPES (Coordena\c c\~ao de Aperfei\c coamento de Pessoal
de N\'\i vel Superior) and FAPESP (Funda\c c\~ao de Amparo \`a
Pesquisa do Estado de S\~ao Paulo) for partially supporting the
WPCF 2006, as well as SBF (the Brazilian Physics Society) for
their assistance  during  the  organization  of  this  meeting.


\end{document}